\documentclass[nofootinbib,aps,superscriptaddress,floatfix,prd]{revtex4}

\usepackage{epstopdf}
\usepackage{bm}
\usepackage{graphicx,epsf,subfig}
\usepackage{amsmath, amsthm, amssymb} 
\usepackage{epsfig}
\usepackage{amsmath,slashed}
\usepackage{footmisc,amssymb,color}

\newcommand{\rig}{\rightarrow}

\newcommand{\be}{\begin{eqnarray*}}
\newcommand{\ee}{\end{eqnarray*}}
\newcommand{\beq}{\begin{equation*}}
\newcommand{\eeq}{\end{equation*}}
\newcommand{\beeq}{\begin{equation}}
\newcommand{\eeeq}{\end{equation}}
\newcommand{\gl}[1]{(\ref{#1})}

\newcommand{\bee}{\begin{eqnarray}}
\newcommand{\eee}{\end{eqnarray}}

\def\lesssim{\mathrel{\raisebox{-.6ex}{$\stackrel{\textstyle<}{\sim}$}}}

\newcommand{\gev}{{\mathrm{~GeV}}}

\newcommand{\eps}{\varepsilon}
\newcommand{\ep}{\eps}

\newcommand{\eg}{{\it e.g.}}
\newcommand{\ie}{{\it i.e.}}
\newcommand{\cf}{{\it cf.}}

\begin{document}

%%%%%%%%%%%%%%%%%%%%%
%%%%%%%%%%%%%%%%%%%%%
\title{Unconstraining the Unhiggs} 
%%%%%%%%%%%%%%%%%%%%%
%%%%%%%%%%%%%%%%%%%%%

%%
\author{Christoph~Englert}
\email{christoph.englert@durham.ac.uk}
\affiliation{Institute for Particle Physics Phenomenology, 
Department of Physics, Durham University, Durham DH1 3LE, UK}
\author{Michael~Spannowsky}
\email{michael.spannowsky@durham.ac.uk}
\affiliation{Institute for Particle Physics Phenomenology, 
Department of Physics, Durham University, Durham DH1 3LE, UK}
\author{David~Stancato}
\email{dastancato@ucdavis.edu}
\affiliation{Department of Physics, University of California, Davis, CA 95616, USA}
\author{John~Terning}
\email{terning@physics.ucdavis.edu}
\affiliation{Department of Physics, University of California, Davis, CA 95616, USA}

%%%%%%%%%%%%%%%%%%%%%
\begin{abstract}
  We investigate whether or not perturbative unitarity is preserved in
  the Unhiggs model for the scattering process of heavy quarks and
  longitudinal gauge bosons $\bar q q \to V_L^+ V_L^-$.  With the
  Yukawa coupling given in the original formulation of the Unhiggs
  model, the model preserves unitarity for Unhiggs scaling dimensions
  $d\leq 1.5$. We examine the LHC phenomenology that is implied by the
  Unhiggs model in this parameter range in detail and discuss to what
  extent the LHC can test $d$ if an excess is measured in the
  phenomenologically clean $ZZ$ channel in the future or if the LHC
  measurement remains consistent with the background. We then make use
  of the AdS/CFT correspondence to derive a new Yukawa coupling that
  is conformally invariant at high energies, and show that with this Yukawa
  coupling the theory is unitary for $1 \leq d < 2$.
\end{abstract}
%%%%%%%%%%%%%%%%%%%%%

%%%%%%%%%%%%%%%%%%%%%
\preprint{IPPP/12/08}
\preprint{DCPT/12/16}
%%%%%%%%%%%%%%%%%%%%%

\maketitle

\section{Introduction}

Georgi \cite{david1,david2}, introduced a new approach to studying
conformal sectors by specifying the two-point functions of scalar
fields with a scaling dimension between one and two. Since the phase
space for these fields resembles the phase space of a fractional
number of particles, Georgi termed them ``unparticles''. Subsequently,
efforts were made to gauge unparticle actions in a consistent way
\cite{david3} so that unparticles could be given Standard Model (SM)
gauge quantum numbers. This also necessitates the introduction of an
IR cutoff for the unparticle sector \cite{david3,david4,david5} so
that there are no new massless modes, which would dramatically alter
low energy phenomenology. Reference \cite{david6} introduced the
Unhiggs as a way to break electroweak symmetry via an unparticle (see
\cite{david8,david9} for work on related ideas). The Unhiggs has the
same gauge structure as the SM Higgs with a scaling dimension $d$ and
an IR cutoff $\mu$. The effects of an Unhiggs on precision electroweak
measurements have been studied in detail \cite{david10,beneke}, and
the model is consistent with the current data. In \cite{david6}, it
was demonstrated that like the SM Higgs, the Unhiggs unitarizes $WW$
scattering and also that the Unhiggs can help ease the Little
Hierarchy Problem. Reference \cite{david7} began to look into
phenomenological aspects of the Unhiggs model, specifically the top
quark decay $t \to W^+b$. Due to the fact that the longitudinal $W$
bosons are affected by the electroweak symmetry breaking sector, the
fraction of emitted longitudinal $W$ bosons is different in the
Unhiggs model than it is in the SM. This information was used, along
with CDF data and upcoming anticipated Large Hadron Collider (LHC)
data, to put current and expected bounds on the $(\mu,d)$ parameter
space in the Unhiggs model.
In this paper we comment on the LHC phenomenology of the Unhiggs model
in the parameter range $d\leq1.5$ and return to a more theoretical
aspect of the Unhiggs model, that of perturbative unitarity for
$d>1.5$.

We examine to what extent the Unhiggs can be constrained, ruled out or
even measured statistically significant at the LHC if $d\leq 1.5$ in
the heavy mass region. This is the mass range where Higgs decays in
the SM are dominated by decays to massive gauge bosons. It is
difficult to get around the current exclusion bounds \cite{zzchannel}
in this channel as a consequence of electroweak symmetry breaking
\cite{Englert:2011us}, which guarantees $\Gamma(H\to ZZ) \sim
m_H^3/m_Z^2$ for $m_H\gtrsim 2m_Z$.  We will see that the Unhiggs
model is an efficient realization of such a hiding mechanism, and from
this point of view the Unhiggs phenomenology in the clean $ZZ\to
4\ell$ final state is especially interesting to reinterpret the
current and future bounds on SM-like production
\cite{ATLASCouncil,atlashiggs,CMSCouncil,cmshiggs,Carmi:2012yp}.

For the parameter choice $d\leq 1.5$ we can expect a valid effective
theory description by the Unhiggs model up to several TeV
\cite{david6}.  We will see that a non-observation of the Higgs boson
can be consistently interpreted in this model.
Increasing $d$ beyond 1.5, we show that the Yukawa coupling in the
original formulation of the Unhiggs \cite{david6}, which explicitly breaks the
conformal symmetry, is the root cause of unitarity violation in
massive quark scattering $\bar q q \to W^+W^-$.  We show analytically
that this ``non-conformal'' Yukawa coupling is not sufficient to
preserve unitarity for values of the scaling dimension $d$ greater
than 1.5.  To preserve unitarity, we need to derive a form of the
Yukawa coupling which respects the conformal symmetry of the model. We
accomplish this by appealing to the AdS/CFT correspondence \cite{adscft},
considering an AdS dual to the Unhiggs model. Once gauged, this
``conformal'' Yukawa coupling will also necessitate the presence of
terms in the Lagrangian with arbitrary numbers of gauge bosons. Taking
these new terms into account, we find that unitarity is indeed
preserved for all values of $d$ in the range $1 \leq d < 2$.

\section{Mixed Yukawa-gauge sector theory and phenomenology}

\subsection{Remarks on perturbativity for $1\leq d \leq 1.5$}
\begin{figure}
  \begin{center}
    \includegraphics[height=4cm]{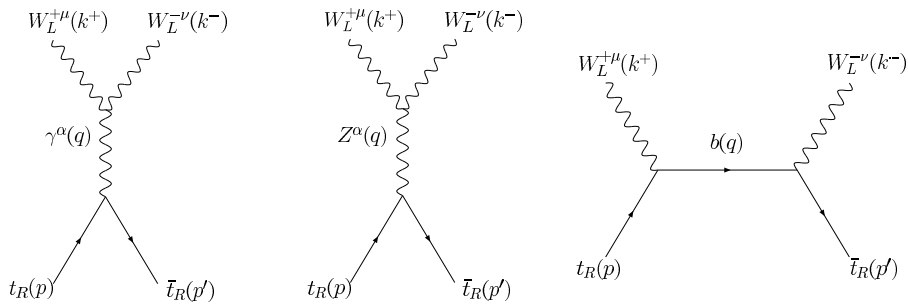}
  \end{center}
  \caption{\label{fig:david1} The three diagrams not involving the
    Unhiggs.}
\end{figure}

\begin{figure}
  \begin{center}
    \includegraphics[height=5cm]{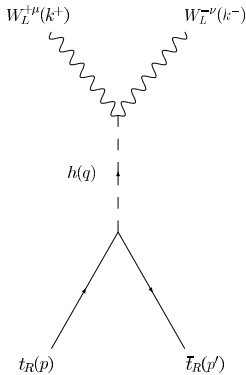}
  \end{center}
  \caption{\label{fig:david2} The $s$-channel Unhiggs exchange
    diagram.}
\end{figure}

We show first that the Yukawa coupling in \cite{david6} which is the
most naive extension of the SM Yukawa coupling to an unparticle, is
sufficient to preserve perturbative unitarity in the scattering
process $\bar t t \to V_L V_L$ ($V=W^\pm,Z$). We will assume the
scattering particles are top quarks, but the same general analysis
follows for the other SM fermions as well. In the SM, there are four
tree level diagrams contributing to $\bar t t \to W_L W_L$. These
diagrams involve exchange of an $s$-channel photon, an $s$-channel $Z$
boson, a $t$-channel bottom quark and an $s$-channel Higgs, and all
have an analog in the Unhiggs model (see figures \ref{fig:david1} and
\ref{fig:david2}). The three diagrams in figure \ref{fig:david1} that
do not contain an Unhiggs have the same amplitude as in the Standard
Model. This is obvious for the diagram with the $t$-channel bottom
exchange because the bottom-top-$W$ vertices as well as the bottom
propagator are completely independent of the EWSB sector.

The $s$-channel photon and $Z$ diagrams also yield the same amplitudes
as in the SM, although it is not as obvious, because not only are
their propagators different in the Unhiggs model, but there are
additional vertices not present in the SM which would seem to
contribute new diagrams to the scattering amplitude. First we will
address the propagators. The photon and $Z$ propagator in the Unhiggs
model are both of the form
\begin{equation}
  \label{eq:prop} 
  \Delta(q^2)={-i\over q^2-M_Z^2}
  \left(g_{\mu\nu} -f(q^2)q^\mu q^\nu \right) 
\end{equation} 
The function $f(q^2)$ is different for the photon and $Z$, and both
are different from their counterparts in the SM. These functions are
given in \cite{david6}, but their actual form is unimportant because
we will show that any term proportional to $q^\mu q^\nu$ in the
propagator disappears in the amplitude for the $s$-channel exchange
diagram. By contracting the second term in the propagator with the
$W^+W^-A$ or $W^+W^-Z$ SM vertex and the polarization vectors, we find
that both amplitudes are proportional to
\begin{equation} 
    {\cal{M}}\supset{\cal{F}}^\mu q_\mu q_\nu\left[ (q+k^+)^\lambda
    g^{\nu\alpha} +(k^--k^+)^{\nu}g^{\alpha\lambda} -
    (q+k^-)^{\alpha}g^{\nu\lambda}\right]
  \ep_\lambda(k^-)\ep_\alpha(k^+)
\end{equation} 
where $k^+(k^-)$ are the momenta of the $W^+(W^-)$ bosons, and $F^\mu$
is a Dirac chain representing the rest of the amplitude. Using $q =
p+p' = (2E,0,0,0)$, where $p,p'$ are the quark momenta
(\cf~Fig.~\ref{fig:david1}), we find that
\begin{align}
  q\cdot (k^--k^+) =& 0\\
  q\cdot \ep(k^-) =& 2E^2/M_W \\
  q \cdot \ep (k^+) =& 2E^2/M_W \\
  (q + k^+) \cdot \ep(k^-) = &4E^2/M_W \\
  (q + k^-) \cdot \ep(k^+) = &4E^2/M_W\,.
\end{align} 
Plugging in, this gives us ${\cal{M}}\sim {E^4/ M_W^2}[ 8+0-8] =0$.
Thus, the part of the propagator proportional to $q_\mu q_\nu$ does
not contribute to the amplitude and therefore the fact that the
functions $f(q^2)$ are different in the Unhiggs model and the SM does
not affect the amplitudes.

Next we must address the fact that there are new vertices in the
Unhiggs model which naively contribute to the $s$-channel photon and
$Z$ amplitudes. Because of the non-local kinetic term in the Unhiggs
model, there are vertices with two Unhiggs and arbitrary numbers of
gauge bosons. Specifically there are $W^+W^-AHH$ and $W^+W^-ZHH$
vertices, and upon taking the Unhiggs to VEVs, these new three gauge
boson vertices contribute to the amplitudes for the $s$-channel
scattering processes. However, by using the method of gauging
non-local Lagrangians given in \cite{david13}, we find that these
vertices contain only terms proportional either to $k_\mu^+$ and/or
$k_\nu^-$. Upon contracting these vertices with the polarization
vectors $\ep_\mu(k^+)$ and $\ep_\nu(k^-)$ the amplitude
vanishes. Therefore, the amplitudes resulting from the three diagrams
shown in figure \ref{fig:david1} all take the same values as in the
SM.

%%%%%%%%%%%%%%%%%%%%%%%%%%%%%%%%%%%%%%%%%%%%%
%%%%%%%%%%%%%%%%%%%%%%%%%%%%%%%%%%%%%%%%%%%%%

%%%%%%%%%%%%%%%%%%%%%%%%%%%%%%%%%%%%%%%%%%%%%
%%%%%%%%%%%%%%%%%%%%%%%%%%%%%%%%%%%%%%%%%%%%%

The results for these amplitudes in the SM are given in
\cite{david11}. The cases where the top and anti-top have opposite
helicities are not of interest because the Unhiggs exchange diagram is
only non-zero when the helicities are either both right-handed or both
left-handed. In other words, in the cases with opposite helicity, the
three non-Higgs diagrams are all that is needed to preserve
unitarity. In \cite{david11} it is indeed explicitly shown that these
three diagrams conspire together to cancel all terms which are
proportional to positive powers of the center of mass energy
$\sqrt{s}$. As for the cases where the helicities of the top and
anti-top are the same, the results are also given in \cite{david11}. We
will write the results for the top and anti-top having right-handed
helicities here for future reference, keeping only the terms which
grow with energy.
\begin{align}
  {\cal{M}}_{\gamma,RR}=& 4\sqrt{2}G_F \left({2\over 3}\right) 
  m_ts^2_w \cos\theta \sqrt{s} \\
  {\cal{M}}_{Z,RR} =& \sqrt{2} G_F m_t \left[ 4 \left( {2\over 3}
    \right) s^2_w - 1\right] \cos\theta \sqrt{s} \\
  {\cal{M}}_{b,RR} =& \sqrt{2} G_F m_t(1 - \cos \theta) \sqrt{s}
\end{align}
where $\theta$ is the angle between the outgoing top quark and
$W_L^-$. Adding these three contributions, we get
\begin{equation}
  \label{eq:david2.6}
  {\cal{M}}_{\gamma+Z+b,RR}= \sqrt{2}G_F m_t\sqrt{s} \,.
\end{equation}
In the Standard Model, the $s$-channel Higgs exchange diagram is given
by
\begin{equation} 
  {\cal{M}}_{H,RR}= -\sqrt{2}G_F m_t\sqrt{s} \,.
\end{equation}
The cases with both particles left-handed is identical other than an
opposite overall sign. We will therefore consider only the
right-handed case for the rest of the paper with no loss of
generality. Clearly in the SM the terms which grow with energy cancel
once the Higgs exchange is included.

%%%%%%%%%%%%%%%%%%%%%%%%%%%%%%%%%%%%%%%%%%%%%%%%
\begin{figure}[t]
  \begin{center}
    \includegraphics[width=0.5\textwidth]{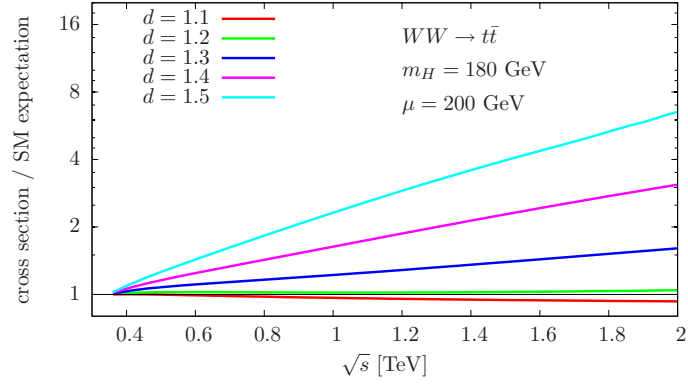}
  \end{center}
  \caption{\label{fig:ttww} The $W^+(m_W)W^-(m_W)\to t\bar t$ cross
    section for different values of the Unhiggs scaling dimensions
    $d$.}
\end{figure}
%%%%%%%%%%%%%%%%%%%%%%%%%%%%%%%%%%%%%%%%%%%%%%%%

\subsubsection*{The Unhiggs exchange diagram}
We now show that the Unhiggs exchange diagram in figure
\ref{fig:david2}, using the naive ``non-conformal'' Yukawa coupling
from \cite{david6}, is not sufficient to preserve unitarity for the
entire range of $d$, $1\leq d < 2$. The Yukawa term in the Lagrangian
in the original paper is given by
\begin{equation} 
  {\cal{L}}_{Y,NC}=-\lambda_t \bar t_R{H^\dagger \over
    \Lambda^{d-1}} \left( \begin{matrix} t \\ b \end{matrix} \right)_L
  + {\hbox{h.c.}} \,.
\end{equation}
where the ``$NC$'' stands for non-conformal. This Yukawa coupling is
identical to that in the SM other than dividing by a power of the UV
cutoff $\Lambda$ which compensates for the fact that the Unhiggs
doublet has dimension $d$ instead of one. This leads to a coupling
between the physical Unhiggs and $\bar t t$ given by
\begin{equation}
  \label{eq:david2.10}
  \Gamma_{Y,NC}={im_t\over v^d}
\end{equation}
with
\begin{equation}
  m_t={\lambda_t v^d\over \sqrt{2}\Lambda^{d-1}} \,.
\end{equation}
The $W^+W^-H$ vertex, when contracted with the polarization vectors,
is given by \cite{david6}
\begin{equation}
  ig^2\Gamma^{+-\mu \nu} \ep_\mu(k^-) \ep_\nu(k^+)
  =i {g^2v^d\over 2} \ep(k^-)\cdot\ep(k^+) {\cal{K}}(0,k^-+k^+)
\end{equation}
where
\begin{equation}
  {\cal{K}}(0,k^-+k^+)=
  - { (\mu^2 - s)^{2-d} - ( \mu^2 )^{2-d}  \over s}
\end{equation}
and the Unhiggs propagator is given by
\begin{equation}
  \Delta_H(s)={-i\over m^{4-2d}-\mu^{4-2d}+(\mu^2-s)^{2-d}}\,.
\end{equation}
Putting all of this together, the amplitude for the $s$-channel
Unhiggs scattering (with helicities unspecified for now) is then given
by
\begin{equation}
  i{\cal{M}}_{H,NC} = ig^2\bar{v}(p')\Gamma_Y u(p)\Delta_H(s)
  \Gamma^{+-\mu\nu}\ep_\mu(k^-)\ep_{\nu}(k^+)
\end{equation}
where $p$ and $p'$ are the momenta of the incoming $t$ and $\bar t$
quarks respectively. After specifying the helicities (both
right-handed) of the fermions and the polarizations (longitudinal) of
the $W$s, we find that in the limit that $s\gg M_W^2 ,m^2_t$
\begin{equation} 
  \label{eq:david2.15}
  {\cal{M}}_{H,NC} =-\sqrt{2}G_Fm_t\sqrt{s} \left[
    {\mu^{4-2d} - (\mu^2 -s)^{2-d} \over \mu^{4-2d} - m^{4-2d} -
      (\mu^2 - s)^{2-d}} \right].
\end{equation}
Finally we add this to the contribution to the amplitude from the
other three diagrams given in Eq. \gl{eq:david2.6} to get
\begin{equation} 
  {\cal{M}}_{NC}= -\sqrt{2}G_Fm_t\sqrt{s}
  {m^{4-2d}\over \mu^{4-2d} - \mu^{4-2d} - (\mu^2 - s)^{2-d}}
\end{equation}
We can now safely take the limit $s \gg \mu^2, m^2$ which yields
\begin{equation}
  {\cal{M}}_{NC}=\sqrt{2}G_Fm_tm^{4-2d}(-1)^{d-2}s^{d-3/2}
\end{equation}
The amplitude is an increasing function of $s$ for $d > 1.5$, with the
consequence that unitarity is not preserved in this process for $d \gtrsim
1.5$ (figure \ref{fig:ttww}).

\subsection{Elements of Unhiggs phenomenology at the LHC for $1\leq d \leq 1.5$}
In this section, we comment on the collider phenomenology that arises
from the modified Yukawa and gauge sector of the Unhiggs model in the
parameter range $1\leq d \lesssim 1.5$.  There are two sources of
potential deviations of the Unhiggs scenario from SM-like production
due to the underlying conformal structure of the symmetry breaking
sector, which can be accessed in production processes of massive
electroweak bosons \cite{david6}:
\begin{enumerate}%[(i)]
\item By eating a continuum Goldstone boson, the phase space of the
  longitudinal $W$ receives a continuum contribution above the
  conformal symmetry breaking scale $\mu$,
  \begin{subequations}
    \label{eq:xsecps}
    \begin{equation}
      d\Phi_W=2\pi
      \theta(q^0)\, \delta(q^2-m_W^2) + \theta(q^0)\,\theta(q^2-\mu^2)\,
      f(q^2)\,, 
    \end{equation} 
    where
    \begin{equation}
      f(q^2)=\frac{-2(2-d)\mu^{2-2d}\sin(\pi
        d)(q^2-\mu^2)^{2-d}}{\mu^{8-4d}+(q^2-\mu^2)^{4-2d}-2\mu^{4-2d}\cos(\pi
        d)(q^2-\mu^2)^{2-d}} \,.
    \end{equation}
  \end{subequations}
\item The Unhiggs, the $W,Z$ propagators and the trilinear $HZZ$ and
  $HWW$ vertices are modified with respect to the SM, yielding, \eg,
 \begin{equation} 
   \label{eq:xsecshape}
   {\sigma^{\rm{Unh}} (gg\rig H \rig VV)\over 
     \sigma^{\rm{SM}}(gg\rig H \rig VV)} \sim \left| {\left(1 -
         {m_H^2\over q^2}\right) }\frac {[(\mu^2 - q^2)^{2 - d} - 
       \mu^{4 - 2d}]}{(\mu^2 - q^2)^{2 - d}-(\mu^2 -
       m_H^2)^{2 - d}}
   \right|^2_{q^2=\sqrt{s}}\,,
 \end{equation}  
 where $\sqrt{s}$ is the partonic center of mass energy as usual. Note
 that for $d=1$ we have $\sigma^{\rm{Unh}}/\sigma^{\rm{SM}}=1$.
\end{enumerate}

\begin{figure}[!t]
  \begin{center}
    \includegraphics[width=0.5\textwidth]{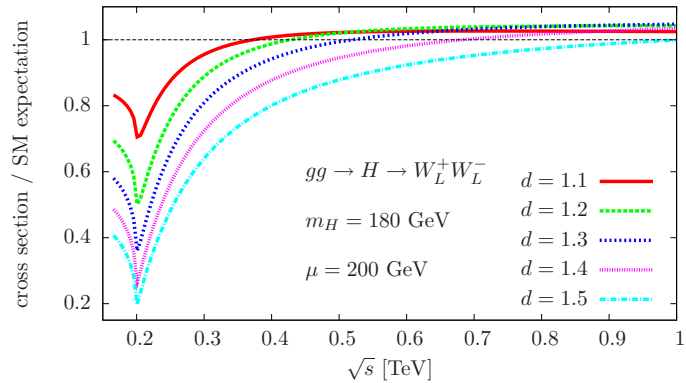}
  \end{center}
  \caption{\label{fig:plot_gg} Comparison of Unhiggs production and
    decay via gluon fusion, $gg\rig H \rig W^+W^-$ for different
    values of the Unhiggs scaling dimension $d\leq 1.5$ for
    identically chosen widths. $\sqrt{s}$ denotes the partonic center
    of mass energy.}
\end{figure}

We include both effects in figure~\ref{fig:plot_gg}, which compares
$gg\rig H \rig W^+_LW^-_L$ in the Unhiggs scenario and the Standard
Model (SM) at the parton level for the typical scales which are probed
at the LHC. We do not include the parton distribution functions at
this point to make the comparison more transparent; comparing the main
production mode contributing to the $pp\rig VV+X$ cross sections at
the LHC for Unhiggs and SM production, the gluon parton distribution
have no phenomenological impact since they are identical. Integrating
over the partonic energy fractions in computing the hadronic cross
section, characteristics of comparison such as the dip for
$\sqrt{s}\sim \mu$ are completely washed out.

The cross section qualitatively follows Eq.~(\ref{eq:xsecshape}) with
the continuum contributions of Eq.~(\ref{eq:xsecps}) ranging at the
percent-level for fully-inclusive production. An important feature of
Unhiggs production $pp\to H$ is that, with respect to the SM, the
cross section decreases for scaling dimensions $d>1$. This is a
consequence of the Unhiggs' gauge interactions
\begin{equation}
\label{eq:switchoff}
  {\cal{L}} \supset H^\dagger \left( D^\mu D_\mu + \mu^2 \right)^{2-d}H \,,
\end{equation}
where $D_\mu$ denotes the familiar $SU(2)_L\times U(1)_Y$ gauge
covariant derivative. Consequently, the gauge interactions of the
Unhiggs can be dialed away by increasing $d$ while keeping the $W$ and
$Z$ masses fixed. This can lead to a qualitatively different
phenomenology compared to SM gluon fusion: the total Unhiggs cross
section can remain small even if large corrections from strong
interactions \cite{gfusion} are taken into account.  At the same time,
however, the Unhiggs branching ratios to massive gauge bosons decrease
and, depending on the Higgs mass region, the Unhiggs can avoid the
current LHC bounds~\cite{atlashiggs,cmshiggs}.

The phenomenology of longitudinal polarizations of the massive $W$s
and $Z$s probes the mechanism of electroweak symmetry breaking and is
generically buried in a much larger transversely-po\-la\-rized $V$
spectrum. The longitudinal polarizations can, in principle, be
extracted by exploiting the $V_L$'s characteristic decay and radiation
pattern. While these effects are well-investigated for weak boson
fusion (WBF) \cite{zeppenfeld} where they have motivated a central jet veto to enhance
longitudinal over transverse polarizations \cite{Bagger:1995mk}, only
recently in Ref.~\cite{Han:2009em} similar concepts have been applied
to final state jet correlations in the context of subjet analyses. In
total, however, getting a handle on the longitudinal part of the
massive weak bosons remains an experimentally as well as theoretically
ambitious task at hadron colliders.

%%%%%%%%%%%%%%%%%%%%%%%%%%%%%%%%%%%%%%%%%%%%%%%%
%%%%%%%%%%%%%%%%%%%%%%%%%%%%%%%%%%%%%%%%%%%%%%%%

Applied to our case, the effect of Eq.~\gl{eq:xsecps} amounts to
measuring a small excesses over the SM expectation in the invariant
$V_LV_T,V_LV_L$ mass tails. Eventually, this becomes a hopeless
endeavor at hadron colliders once theoretical and experimental
systematic uncertainties of both the signal and the background of the
order of a few percent are properly taken into account. Hence, in the
actual experimental analysis, the focus is on reconstructing the
unitarizing resonance mass. To dig out a resonance from a large
continuous background we have to reconstruct the heavy gauge boson
masses from their decay products.  This is achieved by requiring the
$V$ decay product candidates to recombine the $V$ mass within a
certain mass window, which is essentially set by the experiment's
resolution (see Refs.~\cite{prod12,Hackstein:2010wk,Cui:2010km}). This
eventually removes the reducible backgrounds to a large extent, but
also the continuum contribution in Eq.~(\ref{eq:xsecps}) is excluded
from the experimentally observed signal cross section due to the
imposed selection criteria on either the fully-reconstructed invariant
mass distribution or the transverse mass distribution of the
partially-reconstructed final states (\eg~by projecting on events
around the jacobian peak in the transverse cluster mass
\cite{Bagger:1995mk}).

%%%%%%%%%%%%%%%%%%%%%%%%%%%%%%%%%%%%%%%%%%%%%%%%
\begin{figure}[!b]
  \begin{center}
    \includegraphics[width=0.5\textwidth]{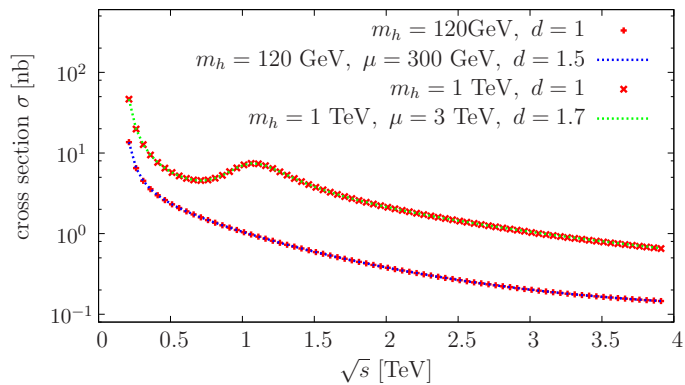}
  \end{center}
  \caption{\label{fig:plot_ww} Comparison of $W^+(m_W)W^-(m_W)\rig
    W^+(m_W)W^-(m_W)$ cross section for Unhiggs production (lines) and
    in the SM (points and crosses). For comparison reasons we plot the
    cross section for identical widths. The cross section is to be
    understood as the double-pole part of the phase space integration
    Eq.~\gl{eq:xsecps}, which reflects the experimental situation
    after reconstructing the final state $W$s.}
\end{figure}
%%%%%%%%%%%%%%%%%%%%%%%%%%%%%%%%%%%%%%%%%%%%%%%%

As a consequence of the unitarization of the longitudinal $VV$
($V=W,Z$) scattering amplitudes as demonstrated in Ref.~\cite{david6},
the phenomenology of WBF processes \cite{Bagger:1995mk,Englert:2008tn}
remains unmodified over the entire parameter range in $(d,\mu,m_H)$,
except for the Higgs width modifications (see below). This directly
follows from the comparison of the unpolarized $WW\rig WW$ cross
section in figure~\ref{fig:plot_ww}, where we plot the cross sections
for identically chosen widths to make the comparison more
transparent. Indeed the Unhiggs width does depend on the chosen values
of $d$ and $\mu$ (see below) but we will not discuss this in the
context of WBF in more detail. Modified Higgs widths and the resulting
modified cross sections with respect to the SM in WBF have been
studied in \eg~Ref.~\cite{modifiedwidth}.

The $WW$ scattering cross section can be straightforwardly generalized
to WBF with leptonic final states in the effective $W$ approximation
\cite{Dawson:1984gx}.  There is no quantitative modification when
abandoning the effective $W$ approximation in favor of the full WBF
matrix element in the Unhiggs scenario for, \eg~$pp\rig 2j
2\ell\slashed{E}_T$. The $W,Z$ propagators in the Unhiggs model show
modifications compared to the SM only in terms proportional to the
four momentum, Eq.~\gl{eq:prop}.  Hence, potential modifications are
parametrically suppressed by contracting the $W,Z$ with light fermion
currents in the full $2\rig 6$ WBF topologies
\cite{Jager:2006zc}. This also accounts for the SM backgrounds to
(Un)Higgs production.
 
Hence, we focus in the following on gluon fusion as production mode
and consider decays $H\rig VV$. This channel can be separated from WBF
by imposing jet vetos. Nonetheless, since WBF phenomenology
remains largely unmodified for $d>1$ we can use the WBF channel to
cross check the gluon fusion results and further constrain the Unhiggs
scenario. Gluon fusion provides the largest production cross section
for Higgs boson measurements at the LHC. Consequently, $gg\to H$ has
received lots of attention from the phenomenology community (see
\eg~Refs.~\cite{gfusion,higgs_review} for an overview).  Along with
the measurement of the Higgs branching ratios to massive gauge bosons,
it provides a channel to study the interplay of the Yukawa and the
gauge sectors\footnote{{\it{E.g.}} by comparing to WBF as a purely
  electroweak production mode, which is also unaltered in our case,
  except for the modified Higgs width.}. This way we can
experimentally infer deviations from the predicted SM rate, which
naturally arise in the Unhiggs model.

As we will see, due to Eq.~\gl{eq:switchoff}, putting stringent bounds
on the Unhiggs parameters $(d,\mu,m_H)$ benefits from a clean
environment and a very good background estimation. Quoting reliable
confidence levels and discovery reaches therefore requires the
inclusion of all contributing experimental uncertainties and $H\to ZZ$
provides a best-case channel to distinguish the SM Higgs from an
Unhiggs. In order to minimize the impact of experimental systematics
on our results we therefore investigate the mass regime $m_H\gtrsim
2m_Z$ with the Higgs decaying fully leptonically via two $Z$s in
detail in this section. This so-called ``gold-plated mode'' offers the
cleanest environment to study the Higgs sector's anatomy, with only
minimal impact of experimental uncertainties \cite{goldplatedmode}.
We note for completeness that Higgs properties can also be
investigated with a boosted semi-hadronic final
state~\cite{Hackstein:2010wk}. The generalization of our results to
fully leptonic $WW$ and semi-hadronic $WW$ and $ZZ$ final states is
straightforward.

%%%%%%%%%%%%%%%%%%%%%%%%%%%%%%%%%%%%%%%%%%%%%%%%
\begin{figure}[!t]
  \begin{center}
    \includegraphics[width=0.5\textwidth]{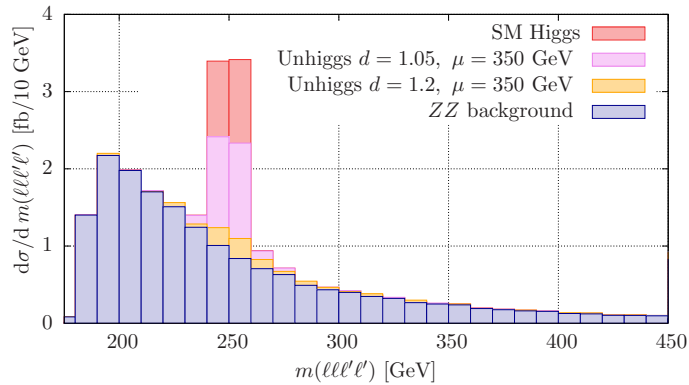}\\[0.3cm]
  \end{center}
  \caption{\label{fig:simul} Invariant four lepton mass in $pp\rig ZZ
    +X \rig 4\ell +X$ at the LHC ($\sqrt{s}=14~{\rm{TeV}}$) for the
    reconstruction requirements quoted in the text. We overlay the
    different signal hypotheses for comparison.}
\end{figure}
%%%%%%%%%%%%%%%%%%%%%%%%%%%%%%%%%%%%%%%%%%%%%%%%

In figure~\ref{fig:simul} we exemplarily compare the $ZZ$ production
(with leptonic decays) for SM Higgs and Unhiggs production for
$m_H=250\gev,~\mu=350\gev,$~ and $d=1.05,1.5$ and $d=1.0$ (the SM
Higgs) For details of the analysis see below. In this mass region, the
LHC collaborations have become sensitive to SM-like cross sections
recently \cite{zzchannel} and we take this as an additional motivation
to discuss the phenomenology that is implied by the Unhiggs scenario
for the eventual high energy LHC run with a target luminosity of
${\cal{L}} \simeq 300~{\rm{fb}}^{-1}$ per experiment in this clean channel.

We adopt the fixed-width complex mass scheme \cite{Denner:1999gp} to
model the Higgs width.  For larger $d$ and $\mu$ in
Eq.~\gl{eq:switchoff}, the Unhiggs decouples rapidly from the final
state gauge boson's phenomenology. At the same time the width
decreases significantly for the chosen physical Unhiggs mass. With the
mass of the $W$ boson in the Unhiggs model \cite{david6} the width
scales with respect to the SM Higgs boson width
\begin{equation}
\label{eq:width}
   \frac{\Gamma^{\rm{Unh}}}{\Gamma^{\rm{SM}}}
    \simeq {(\mu^2)^{d-1}\over 2-d}\left( 
    \frac{(\mu^2)^{2-d}-(\mu^2-m_H^2)^{2-d}}{m_H^2} \right)^2  
    {-\pi {\cal{A}}_d \over 2\pi \sin( \pi d ) } {(  \mu^2-m_H^2  )^{d-1}\over 2-d}\,,
\end{equation}
where
\begin{equation}    
    {\cal{A}}_d={16\pi^{5/2}\over (2\pi)^{2d}} {\Gamma(d+1/2)\over \Gamma(d-1) \Gamma(2d)}\,.
\end{equation}
Note that the Unhiggs width has mass dimension one for arbitrary
values $1\leq d< 2$, and we recover
${\Gamma^{\rm{Unh}}}/{\Gamma^{\rm{SM}}}=1$ upon taking the limit $d\to
1$.

In fact, this is a very good approximation because we have $m_H< \mu$
in such a way that the $W_L$ and $Z_L$ boson can not access the
continuum phase $q^2\geq\mu^2$ in Eq.~\gl{eq:xsecps}. Indeed a large
$\mu$ for fixed $d$ is preferred by electroweak precision constraints
\cite{beneke}, however the phenomenology is largely independent of
$\mu$ Eqs.\gl{eq:xsecps},\gl{eq:xsecshape}, \gl{eq:width}. For the
computation of figure~\ref{fig:plot_gg} we found that the continuum
leaves minor modifications in only the longitudinal
components. Therefore, the above approximation should be sufficiently
good to also govern the phenomenology for more general parameter
choices if the $WW$ and $ZZ$ final states are the dominant Unhiggs
decay channels. Furthermore, the total Higgs width is
phenomenologically dominated by detector effects, and, hence,
measurements of the absolute Higgs width are extremely involved at
hadron colliders. In the clean $H\to ZZ \to 4\ell$, however the width
is dominated by the physical one \cite{zzchannel}. We note for
completeness, that for smaller Unhiggs masses decays to fermions
dominate, but also loop-induced decays can alter the Higgs
phenomenology, especially for $d>1.5$ when a modified fermion sector
is realized. We leave a detailed investigation of this direction to
future work.

For the purpose of this section, we are predominantly interested in
assessing how and if the Unhiggs can be discriminated from the SM
Higgs or the background. We adopt the viewpoint that $d$ is a small
quantity. This implies physics largely consistent with the SM to our
current knowledge by construction.  Consequently the branch cut of the
Unhiggs phase space is subleading and the phenomenology is mainly
affected by the modified Higgs resonance.  We do not include
signal-background interference effects with continuum $gg\to WW$
production \cite{Campbell:2011cu}, which will have an impact in high
precision studies but will be of comparable size in both scenarios. We
produce signal events with a modified version of {\sc{MadEvent}}
\cite{Alwall:2007st} interfaced with
{\sc{Herwig++}}~\cite{Gieseke:2011na} and background events with
{\sc{Sherpa}}~\cite{Gleisberg:2008ta}.  We require a minimum
transverse momentum for four isolated leptons $p_T^{\ell}\geq
20~{\rm{GeV}}$ and a dilepton separation in the azimuthal
angle-pseudorapidity plane $R_{\ell\ell'}\geq
\sqrt{\Delta\eta_{\ell\ell'}^2+\Delta\phi_{\ell\ell'}}\geq 0.2$.
Throughout, we model finite detector resolution effects for leptons
with a gaussian smearing by about
\begin{equation} 
{\Delta E \over E} = 0.02
\end{equation}
around the central true MC value, where $E$ is the lepton's energy
\cite{Armstrong:1994it}.  With an isolated lepton we mean an
identified lepton in the electromagnetic calorimeter $|\eta|\leq 2.5 $
with a hadronic energy deposit of smaller than 10\% of the leptons's
transverse momentum in a cone of size $0.3$ around the lepton. The
leptons are required to reconstruct the $Z$ mass within a
$10~{\rm{GeV}}$ window with same flavor and opposite charge.

In the actual experimental analysis the normalization of the
distributions in the theoretically challenged $pp\rig VV$ production
channel \cite{vv} is preferably extracted from data in control regions
\cite{zzchannel,controlxsec}. The observed signal cross section is
then a function of the background extrapolation. In order to be more
realistic when testing the different hypothesis we conservatively
account for higher order QCD corrections by including a total
$K$-factor $K=2$ for both signal and background\footnote{It is know
  from the literature that $pp \to VV$ \cite{vv} receives large
  corrections ${\cal{O}}(50\%)$ at NLO QCD. NLO corrections to $pp \to
  VV$+jet \cite{vvjet}, which is a part of the NNLO $VV$ production
  cross section is again large $\simeq 50\%$. The higher order
  corrections to Higgs production are also sizable \cite{gfusion}. In
  total higher order corrections have significant impact on our
  search.}

We can already draw a few quantitative conclusion from
figure~\ref{fig:simul}.  Measuring a SM Higgs-like excess in the
considered mass range allows us to tightly constrain the $(d,\mu,m_H)$
parameter range already with low statistics. The smaller width at a
lower physical Unhiggs production cross section leaves a narrow
resonance which is washed out by detector and mass resolution
effects. A precise determination of $d>1$ for fixed $\mu$ and a
statistically established resonance requires larger integrated
luminosities.  We discuss how this is quantitatively reflected in the
full statistical analysis in the next section.

\subsubsection*{Hypothesis testing: Towards exclusion or discovery}
To study these implications on a quantitative level we perform a
log-likelihood shape comparison of the various hypotheses that are
depicted figure~\ref{fig:simul}. This is a standard method in the
post-LEP era \cite{lepera} which allows to compute confidence levels
in a modified frequentist approach \cite{Read:2002hq}.

We define the log-likelihood ratio comparing the SM (null)hypothesis
with the Unhiggs (alternative) hypothesis
\begin{equation}
  \label{eq:loglike2}
  {\cal{Q}}
  = -2\log \frac{L( \text{data}\, |\, {\text{Unhiggs~+~background}} ) } 
  { L( \text{data}\,| \, {\text{SM~Higgs~+~background}} ) }\,.
\end{equation}
$L$ denotes the Poissonian likelihood,~\ie 
\begin{equation}
L( \text{data}\, |\,  {\text{Unhiggs~+~background}}) =
 \frac{N^{n} e^{-N}}{n!}\,,
\end{equation}
where $N=(\sigma^{\rm{Unh}}+\sigma^{\rm{bkg}}){\cal{L}}$ is the number
of expected events at a given luminosity and $n$ is the number of
actually observed events in the Unhiggs model.  Due to the additivity
of the logarithm, the extension to differential distributions is
straightforward. In the described scenario the log-likelihood picks up
sensitivity from mostly around the Higgs peak, giving effectively rise
to the comparison of counting experiments integrated over a few
bins. Potential width-induced modifications of the phenomenology
modulo detector and resolution effects are evidently incorporated in
this approach.

According to the Neyman-Pearson lemma \cite{neyman}
Eq.~\gl{eq:loglike2} is the uniformly most powerful test statistic to
discriminate between our two hypotheses (see
Refs.~\cite{junk,phenoapp} for further details and phenomenological
applications). We sample the two hypotheses with randomly generated
pseudodata, yielding probability density functions (pdfs)
${{Q}}^{\rm{Unh}}({\cal{Q}})$ and ${{Q}}^{\rm{SM}}({\cal{Q}})$, for
the Unhiggs and the SM distribution, respectively. From these pdfs we
derive the confidence levels for exclusion and discovery.
One main advantage of the log-likelihood hypothesis test is that there
is a well-defined statistical notion for small luminosities which
imply only a few bin counts. As already few bin entries can give rise
to strong exclusion due to the large cross section deviations between
the different hypotheses that are depicted in figure~\ref{fig:simul},
this method yields reliable results, as opposed to significances $\sim
{\hbox{signal}}/\sqrt{\hbox{background}}$, which are valid for large
statistics.

%%%%%%%%%%%%%%%%%%%%%%%%%%%%%%%%%%%%%%%%%%%%%%%%
\begin{figure}[!t]
  \begin{center}
    \includegraphics[width=0.42\textwidth]{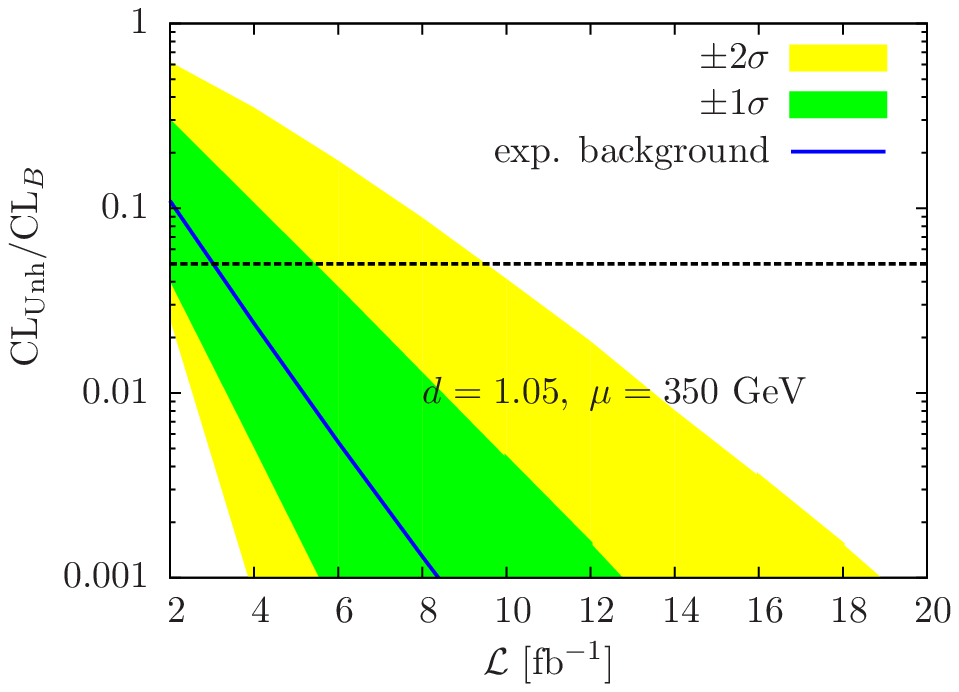}
    \hspace{2cm}
    \includegraphics[width=0.42\textwidth]{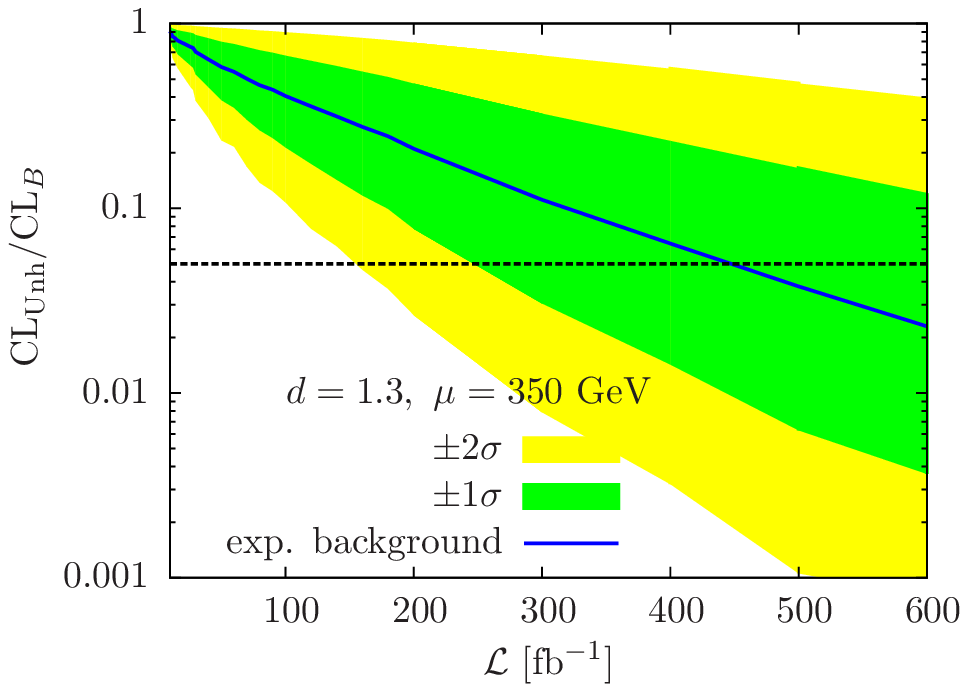} \\[0.3cm]
  \end{center}
  \caption{\label{fig:backcls}Expected exclusion limit for the two
    Unhiggs benchmark points based on the given the background-only
    hypothesis for $\sqrt{s}=14~{\rm{TeV}}$ at the LHC.}
\end{figure}
%%%%%%%%%%%%%%%%%%%%%%%%%%%%%%%%%%%%%%%%%%%%%%%%

%%%%%%%%%%%%%%%%%%%%%%%%%%%%%%%%%%%%%%%%%%%%%%%%
\begin{figure}[!t]
  \begin{center}
    \includegraphics[width=0.42\textwidth]{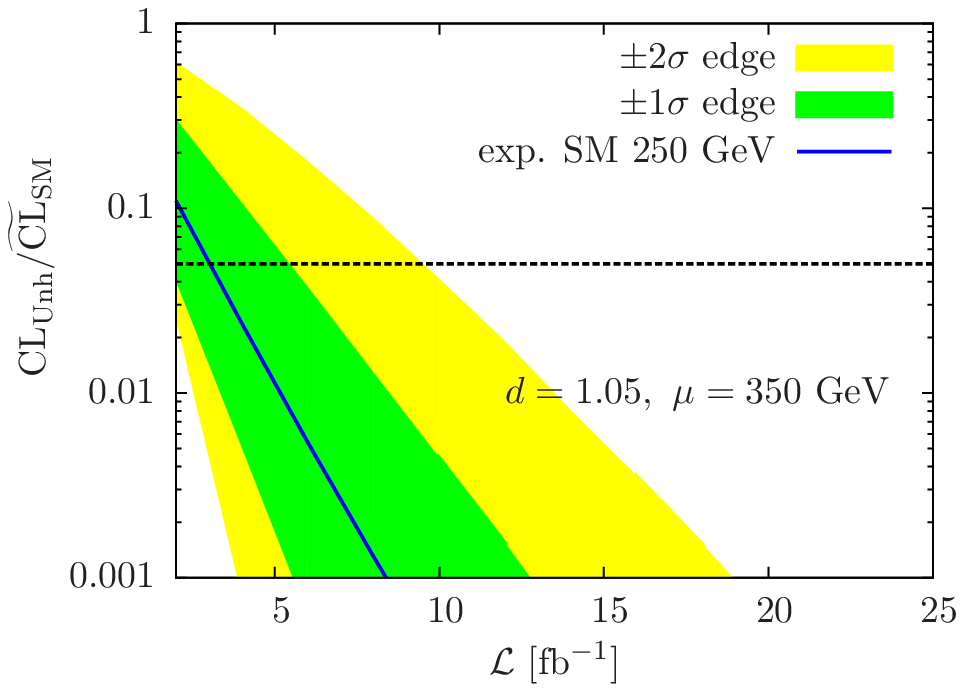}
    \hspace{2cm}
    \includegraphics[width=0.42\textwidth]{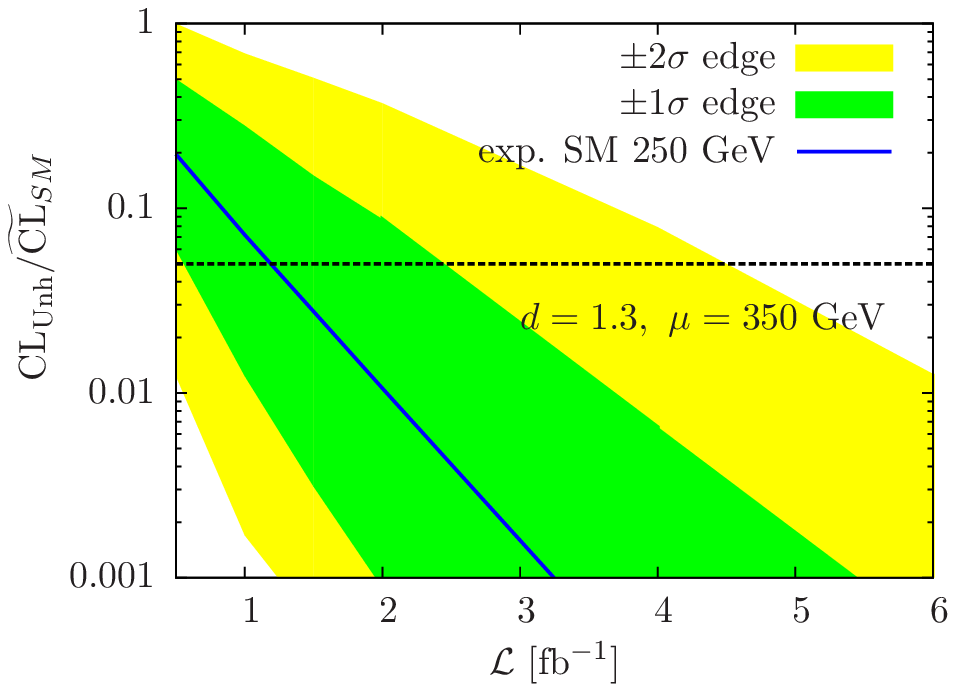} \\[0.3cm]
  \end{center}
  \caption{\label{fig:smcls}Expected exclusion limit for the two
    Unhiggs benchmark points based on the given a SM-like Higgs excess
    null hypothesis for $\sqrt{s}=14~{\rm{TeV}}$ at the LHC.}
\end{figure}
%%%%%%%%%%%%%%%%%%%%%%%%%%%%%%%%%%%%%%%%%%%%%%%%

Exclusion limits are conventionally expressed 
by computing the ${\rm{CL}}_S$ ratio \cite{Read:2002hq}, \ie
\begin{equation}
  {\rm{CL}}_S=\frac{ {\rm{CL}}_{\rm{Unh}} }{
    \widetilde{\rm{CL}}_{\rm{SM}} }\,.  
\end{equation}
Exemplary ${\rm{CL}}_S$ distributions as functions of the integrated
luminosity are shown in figures~\ref{fig:backcls} and \ref{fig:smcls}
for $d=1.05,~1.3$, $m_H=250~\gev$ and $\mu=350~\gev$.  We also include
the ${\rm{CL}}_S$ contours for the $\pm 1\sigma$ and $\pm 2\sigma$
edges of the respective null hypothesis as colored bands. These
parameter choices are motivated by their preferred consistency with
the SM Higgs+background and the background-only hypothesis,
respectively.

The physics community refers to the parameter interval resulting from
${\rm{CL}}_S\leq 0.05$ as ``excluded at the 95\% confidence level'' by
convention \cite{Read:2002hq}; for ${\rm{CL}}_S\leq 0.05$ the Unhiggs
``false exclusion''
\begin{equation}
  \label{eq:clun}
   {\rm{CL}}_{\rm{Unh}} = \int^{\infty}_{
    \left\langle{\cal{Q}}\right\rangle^{\rm{SM}}} d{\cal{Q}}\;
  {{Q^{\rm{Unh}} }}({\cal{Q}})
\end{equation} 
is smaller than 5\% of the potential SM exclusion confidence level
\begin{equation} 
\label{eq:clsm}
 \widetilde {\rm{CL}}_{\rm{SM}} =
  \int^{\infty}_{ \left\langle{\cal{Q}}\right\rangle^{\rm{SM}}}
  d{\cal{Q}}\; {{Q^{\rm{SM}} }}({\cal{Q}})\,,  
\end{equation} 
where $\left\langle{\cal{Q}}\right\rangle$ denotes the median of the
respective sampled pdf. figure~\ref{fig:backcls} displays confidence
levels of Unhiggs+background vs background-only, applying the
identical approach. The expected alternative hypotheses'
${\rm{CL}}_S=0.5$ follows from Eqs.~\gl{eq:clun} and \gl{eq:clsm},
accordingly. 

Figures~\ref{fig:backcls} and ~\ref{fig:smcls} deserve some comments.
The central blue line gives ${\rm{CL}}_S$ as a function of the
integrated luminosity, taking the Monte Carlo-generated data of
figure~\ref{fig:simul} for the respective null hypothesis at face
value (\ie~the expected value of ${\rm{CL}}_S$). The shaded bands
amount to the ${\rm{CL}}_S$ values within $\pm 1\sigma$ and $\pm
2\sigma$ fluctuations of the background distribution at the given
integrated luminosity. If the null hypothesis is not disfavored, the
observed ${\rm{CL}}_S$ value should fall within the shaded bands (see
Ref. \cite{atlashiggs,cmshiggs} for the ${\rm{CL}}_S$ results of
current SM Higgs searches.). Since ${\rm{CL}}_S$ is a ratio of
confidence levels rather then a confidence level itself, its
interpretation does not directly relate to the usual language of a
$5\sigma$ discovery or exclusion. Instead it gives a conservative
interpretation of the measured data, which is protected against
insufficient statistics or insufficiently known backgrounds entering
the measured likelihood ratio \cite{Read:2002hq}.

%%%%%%%%%%%%%%%%%%%%%%%%%%%%%%%%%%%%%%%%%%%%%%%%
\begin{figure}[!t]
  \begin{center}
    \includegraphics[width=0.5\textwidth]{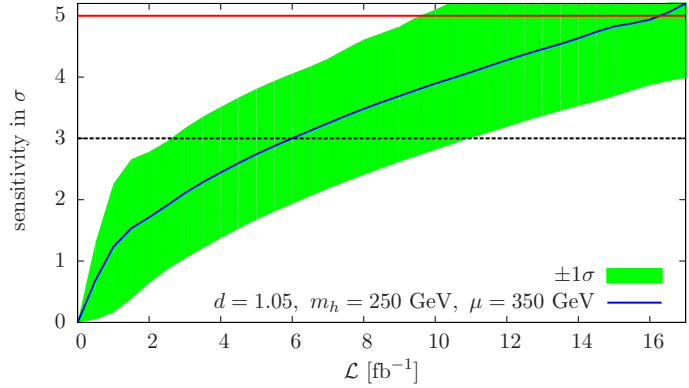}
  \end{center}
  \caption{\label{fig:backdiscov} Statistical significance of
    Unhiggs+background vs background-only as a function of the
    integrated luminosity for $d=1.05,~m_H=250\gev, \mu=350\gev$.}
\end{figure}
%%%%%%%%%%%%%%%%%%%%%%%%%%%%%%%%%%%%%%%%%%%%%%%%

To estimate the luminosity at which we can expect the described
Unhiggs scenarios to be eventually measured statistically significant,
we identify the $p$ value (which relates to the actually measured
value ${\cal{Q}}$ by the experiment) with the median of the
${\text{Unhiggs+background}}$ hypothesis
$\left\langle{\cal{Q}}\right\rangle^{\rm{Unh}}$. This amounts to a
situation when the experiment's outcome exactly follows the Unhiggs
median, thus giving rise to a lower luminosity bound for discovery.
We compute significances from the probability of the background-only
hypothesis to yield a result compatible with the Unhiggs hypothesis by
statistical fluctuations,~\ie~we compute the ``false discovery rate''
\begin{equation}
  \label{eq:CLb}
  1-{\rm{CL}}_{{B}}=
  \int_{-\infty}^{ \left\langle{\cal{Q}}\right\rangle^{\rm{Unh}}} 
  d{\cal{Q}}\; {{Q^{{B}} }}({\cal{Q}})\,.
\end{equation}
A value of $1-{\rm{CL}}_{B}< 2.87\times 10^{-7}$ corresponds to
the familiar notion of a discovery with a significance of larger than
$5\sigma$. In figure~\ref{fig:backdiscov}, we also show the $\pm
1\sigma$ uncertainty band that arises from our computation of
confidence levels.

\bigskip

Figures~\ref{fig:backcls}, \ref{fig:smcls} and \ref{fig:backdiscov}
quantitatively confirm the statements at the end of the previous
section. In general, for $d<1.5$ the increasingly narrow resonance and
the naive suppression of the $pp \to H$ cross section due to
Eq.~\gl{eq:switchoff} leave a generically smaller cross section than
expected in the SM. Consequently, the current exclusion bounds from
the 7 TeV LHC data sample can be avoided. Also, if a SM-like Higgs
excess is measured statistically significant in the near future,
severe constraints on the combination of $(d,\mu)$ follow immediately.

Eventually abandoning the Unhiggs hypothesis in favor of the SM
becomes a question of large integrated luminosity, and, more
importantly, experimental systematics. Since for small $d> 1$ the
Unhiggs quickly approaches the SM, the small differences in the cross
section are difficult to access experimentally, given the residual
uncertainty of the background. In this sense the considered channel is
indeed a best-case scenario. Further discriminating the Unhiggs from
the SM Higgs requires a detailed spectroscopy of the measured excess,
which is performed best at a future linear $e^+e^-$ collider
\cite{Djouadi:2007ik}.  From figure~\ref{fig:smcls} we conclude that
the LHC can potentially probe $d\lesssim 1.3$ level with its target
luminosity. This is also the luminosity at which WBF provides an
additional excellent search channel (see, \eg, \cite{Englert:2008tn}
for a detailed signal and background analysis in the context of the
SM), where the (Un)Higgs resonance is established at the $5\sigma$
level.

\section{The ``conformal'' Yukawa coupling: Using the AdS/CFT
  Correspondence}
In the longitudinal $W$ boson scattering process \cite{david6}, the
only interactions involving the Unhiggs were gauge interactions, which
preserve the conformal invariance at high energies. With the conformal
invariance intact, unitarity is manifest. The root of the unitarity
problem with the fermionic scattering process is that with the
original Yukawa coupling we have explicitly broken the conformal
invariance by introducing the cutoff $\Lambda$. This was the simplest
coupling to introduce, but unlike the case of the gauge-Unhiggs
interactions, it does not preserve the conformal nature of the theory,
which causes unitarity to fail. In the case of the gauge interactions,
there is a simple guiding principle which allows us to exactly
determine couplings that preserve the conformal nature of the
theory. The Unhiggs kinetic term in the action is completely
determined by conformal invariance, and simply gauging this action
yields gauge interactions which respect this invariance. Similarly, we
expect that if we can find a Yukawa coupling which respects the
conformal invariance, the $\bar t t \to W^+_L W^-_L$ process will also
be unitary. Unlike the case of the gauge interactions, which simply
involved gauging the conformally invariant kinetic term, we will have
to make use of the AdS/CFT correspondence between fields in AdS$_5$
and 4D unparticle fields to find a Yukawa term which is conformal at
high energies, as in \cite{david5}. In this section, we detail the
derivation of this ``conformal'' Yukawa coupling and show that unlike
the naive SM-like case above, it is a non-local derivative coupling, a
fact which is crucial for preserving unitarity.

We will use the conformally flat AdS metric
\begin{equation}
  ds^2={R^2\over z^2}\left(\eta_{\mu\nu}dx^\mu dx^\nu - dz^2\right)
\end{equation}
with a brane located at $z =\ep$, which is equivalent to a UV cutoff
$\Lambda = 1/\ep$, and the space extending to infinity. We will take
the limit $\ep\to 0$ at the end of the calculation. The field content
in our model consists of a scalar field and two Dirac fermions. The
scalar field on the boundary will turn out to be the 4D unparticle
field. We aim to calculate the Yukawa coupling between the unparticle
and two fermion zero modes. The scalar action is
\begin{equation}
  \label{eq:phi}
  S_\phi={1\over 2}\int dx\, dz\, \left({R\over z}\right)^3
  \left(\partial_M\phi \partial^M\phi - {m^2R^2\over z^2}\phi^2\right)
\end{equation}
In the fermion sector, we introduce two Dirac fermions $\Psi_L$ and
$\Psi_R$ written in terms of Weyl fermions as follows:
\begin{equation}
  \Psi_L=\left(
    \begin{matrix}
      \chi_L \\ \bar \psi_L
    \end{matrix}
  \right)
\end{equation}
and
\begin{equation}
  \Psi_R=\left(
    \begin{matrix}
      \chi_R \\ \bar \psi_R
    \end{matrix}
  \right) \,.
\end{equation}
The fermion action is given by
\begin{multline}
  \label{eq:fermion}
  S_f = \int d^4x\, dz \left({R\over z}\right)^4 \bigg[
  -i\bar\chi_{L/R}\bar\sigma^\mu\partial_\mu\chi_{L/R}
  -i\psi_{L/R}\sigma^\mu\partial_\mu\bar\psi_{L/R}
  +{1\over 2} \left( \psi_{L/R} \overleftrightarrow{\partial_z}
    \chi_{L/R} - \bar\chi_{L/R} \overleftrightarrow{\partial_z}
    \bar \psi_{L/R} \right)\\
  +{c_{L/R} + m_{L/R} z \over z} (\psi_{L/R}\chi_{L/R} +
  \bar{\chi}_{L/R} \bar\psi_{L/R}) \bigg]\,,
\end{multline}
where $c_{L/R}$ are the bulk masses and $m_{L/R}$ are the coefficients
of the $z$-dependent mass terms, which will allow a subset of the
fermions to have zero modes. The preceding fermion action is designed
to make sure that we get two Weyl fermions which have zero modes in
addition to continuum modes \cite{david12}. Finally, the Yukawa
interaction between the scalar and fermions is given by
\begin{equation}
  \label{eq:yaction}
  S_Y=\int d^4x\, dz \left( {R\over z} \right)^5 {\lambda_5\over 2}
  \phi (\chi_L\psi_R + \chi_R\psi_L ) + {\rm{h.c.}}
\end{equation}
where $\lambda_5$ is a Yukawa coupling of mass dimension $-1/2$.

\subsection{Analysis of the scalar action}
First, we analyze the scalar action in Eq. \gl{eq:phi}. Using
$\partial_M \phi\partial^M\phi=\partial_\mu \partial^\mu\phi
- \partial_z\phi\partial_z \phi$, we find that upon variation of the
action the equation of motion for $\phi$ is
\begin{equation}
  \partial_\mu\partial^\mu \phi - \partial_z^2\phi
  +{3\over z}\partial_z \phi+ {m^2R^2\over z^2}\phi
  =0\,.
\end{equation}
Going to momentum space, our differential equation for $\phi$ (with
4-momentum $p$) becomes
\begin{equation}
  \label{eq:eqm}
  \partial^2_z\phi(p,z)-{3\over z}\partial_z\phi(p,z)+p^2
  \phi(p,z)- {m^2R^2\over z^2} \phi(p,z)=0\,.
\end{equation}
The solution to this equation is
\begin{equation}
  \phi(p,z)=Az^2[aJ_\nu(pz) + bY_\nu(pz)]\,,
\end{equation}
where $\nu=\sqrt{4+m^2R^2}$ and $p\equiv\sqrt{p^2}$.  We will see that
$\nu$ relates to the scaling dimension of the unparticle via the
relation $d_2=2-\nu$.  The values $a$ and $b$ are determined by the
boundary conditions in the IR, $z\to \infty$, and we will leave them
undetermined, while the normalization constant $A$ is determined by
the boundary conditions (BCs) on the UV brane \cite{david5}. We will
impose the UV boundary condition $\phi(p,\ep)=\phi_0(p)$. This
determines $A$, and the solution to $\phi(p, z)$ becomes
\begin{equation}
  \label{eq:ratio}
  \phi(p,z) = \phi_0(p) \left({z\over \ep}\right)^2
  {aJ_\nu(pz) + b Y_\nu(pz) \over aJ_\nu(p\ep) + b Y_\nu(p\ep) }\,.
\end{equation}
Plugging Eq. \gl{eq:eqm} back into the scalar action and integrating
by parts, we find that the integral over $z$ reduces to a pure
boundary term on the UV brane:
\begin{equation}
  S_\phi= {1\over 2} \int {d^4 p \over (2\pi)^4}
  \left[-\left({R\over z} \right)^3 \phi(-p,z)
    \partial_z\phi(p,z)\right]\Bigg|_\ep\,.
\end{equation}
By definition, $\phi(-p,\ep)=\phi_0(-p)$, so we must calculate only
the term $\partial_z \phi(p, z)|_\ep$. Also, we are interested in the
range of scaling dimensions $1 \leq d_S < 2$, which means that $\nu$
is constrained to the range $0 < \nu \leq 1$. For these values of
$\nu$, $Y_\nu(p\ep) \gg J_\nu(p\ep)$ as $\ep\to 0$. Therefore we will
only need to keep the second term from the denominator of
Eq.~\gl{eq:ratio}. Using these facts, and using the series expansions
for the Bessel functions, we find that the action can be separated
into a local term and a non-local term. The local term is given by
\begin{equation}
  S_{\phi,(local)}={1\over 2} \int {d^4 p\over (2\pi)^4} \phi_0(-p)\phi_0(p) \left(
    {R\over \ep} \right)^3\left[ {-(2-\nu)\over \ep} + {\cal{O}}(\ep^0) \right]\,.
\end{equation}
This local term can be removed by UV brane counterterms, and we will
not consider them further \cite{david5}. The non-local term is given
by
\begin{equation}
  S_{\phi,(non-local)}={1\over 2} \int {d^4p \over (2\pi)^4}
  \phi_0(-p)\phi(p) \left({R\over \ep}\right)^3 \left[ {a\over b}
    {\pi \ep^{2\nu-1} \over 2^{2\nu}[\Gamma(\nu)]^2} p^{2\nu} + {\cal{O}}(\ep^{2\nu})
  \right]\,.
\end{equation}
If we rescale $\phi_0$ so that
\begin{equation}
  \phi'_0(p)=\phi_0(p)\left( {R\over \ep} \right)^{3\over 2} 
  \sqrt{a\over b}	{\sqrt{\pi} \ep^{\nu-{1\over 2}}\over 2^{\nu}
    \Gamma(\nu)}
\end{equation}
we find that the non-local part of the action (in the $\ep\to 0$
limit) is
\begin{equation}
  S_{\phi,(non-local)}={1\over 2} \int {d^4p \over (2\pi)^4}
  \phi'_0(-p)p^{2\nu}\phi'_0(p) \,.
\end{equation}
This action is exactly the action for a 4D scalar unparticle $\phi'_0$
with scaling dimension $d_S =2-\nu$.

\subsection{Analysis of the fermion action}
We now proceed to solve for the 5D fermion wavefunctions from the
action given in Eq.  \gl{eq:fermion}. From the variation of $\bar
\chi_{L/R}$ we find
\begin{equation}
  i\bar \sigma^\mu \partial_\mu \chi_{L/R} + \partial_z \bar \psi_{L/R}
  - \left({2+{c_{L/R}}\over z} \right)\bar \psi_{L/R} - m_{L/R}\bar\psi_{L/R}
  =0
\end{equation}
and from the variation of $\psi_{L/R}$ we find
\begin{equation}
  i\sigma^\mu \partial_\mu \bar \psi _{L/R}  - \partial_z
  \chi_{L/R} +
  \left( {2-c_{L/R} \over z} \right) \chi_{L/R} - m_{L/R}\chi_{L/R}=0\,.
\end{equation}
Next we separate the 5D fields as follows:
\begin{align}
\chi_{L/R}(x, z) &= \chi_{0,L/R}(x)g_{L/R}(z) \\
\bar\psi_{L/R}(x, z) &= \bar\psi_{0,L/R}(x)h_{L/R}(z)\,.
\end{align}
Using the 4D Dirac equations
\begin{align}
  i\bar \sigma^\mu\partial_\mu\chi_{0,L/R}&=-p\bar \psi_{0,L/R}\\
  i \sigma^\mu\partial_\mu\bar \psi_{0,L/R}&=-p \chi_{0,L/R}
\end{align}
along with the fact that we are interested in zero modes $(p = 0)$, we
find the following equations for $g_{0,L/R}$ and $h_{0,L/R}$:
\begin{align}
  \partial_z g_{0,L/R} -\left( {2-c_{L/R} \over z } \right) g_{0,L/R} + m_{L/R}g_{0,L/R}& =0 \\
  \partial_z h_{0,L/R} -\left( {2+c_{L/R} \over z } \right) h_{0,L/R} - m_{L/R}h_{0,L/R}& =0 \,.
\end{align}
Solving for the zero mode wavefunctions, we get
\begin{align}
  g_{0,L/R}(z)&=B_0 \left( {z\over R} \right)^{2-c_{L/R}}e^{-m_{L/R}z} \\
  h_{0,L/R}(z)&=C_0 \left( {z\over R} \right)^{2+c_{L/R}}e^{m_{L/R}z}
\end{align}
where $B_0$ and $C_0$ are normalization constants with mass dimension
$1/2$. From the form of the zero mode solutions, we see that
$\chi_{L/R}$ has a normalizable zero mode solution only for $m_{L/R} >
0$ while $\psi_{L/R}$ has a normalizable zero mode solution only for
$m_{L/R} < 0$. We will choose $m_L > 0$ and $m_R < 0$, so that
$\chi_L$ and $\psi_R$ have zero modes, whereas $\chi_R$ and $\psi_L$
do not. The fermions will also have continuum modes (and possibly
discrete modes with non-zero masses, depending on the values of
$c_{L/R}$, see \cite{david12}) but we will only be interested in the
zero modes for now. With the restriction that $m_L > 0$ and $m_R < 0$,
the normalization constants are
\begin{align}
  \label{eq:1a}
  B_0 &=(2m_L R)^{-c_L} (2m_L)^{1\over 2} [ \Gamma(1-2c_L ,2m_L \ep)]^{-{1\over 2}}\\
  C_0 &=(-2m_L R)^{c_R} (-2m_R)^{1\over 2} [ \Gamma(1+2c_R ,-2m_R \ep)]^{-{1\over 2}}
  \,.
\end{align}
 For convenience, we define dimensionless normalization constants as
follows:
\begin{align}
  B_0 &=b_0R^{-{1\over 2}}\\ 
  \label{eq:2}
  C_0 &=c_0R^{-{1\over 2}}\,.
\end{align}

\subsection{Analysis of the Yukawa coupling}
We are now ready to calculate the form of the 4D Yukawa coupling
between the scalar unparticle and the fermion zero modes. After
Fourier transforming the 4D fermion fields, the first term in the
Yukawa action (Eq. \gl{eq:yaction}) becomes
\begin{equation}
  S_Y = \int {d^4 q \over (2\pi)^4} {d^4 k \over (2\pi)^4} {d^4 p
    \over (2\pi)^4}
  \delta^4(p + q + k)\, dz\, \phi_0(p, z) \chi_{0,L}(k)\psi_{0,R}(q)
  {\lambda_5\over \sqrt{2} } \left( {R\over z} \right)^5
  g_{0,L}(z)h_{0,R}(z)\,.
\end{equation}
Plugging in $\phi_0(p,z)$, $g_{0,L}(z)$ and $h_{0,R}(z)$ as well as
substituting $\phi'_0(p)$ for $\phi_0(p)$, we find that the Yukawa
coupling $\Gamma_Y$ between the 4D fields $\phi'_0(p)$,
$\chi_{0,L}(k)$ and $\psi_{0,R}(q)$ is given by
\begin{equation}
  \Gamma_Y=\int dz {\lambda_5\over \sqrt{2}} \left({\ep\over R} \right)^{3\over 2}
  \sqrt{b\over a} {2^\nu \Gamma(\nu)\over \sqrt{\nu}} {\ep^{1\over 2}\over \ep^\nu}
  \left( {z\over \ep} \right)^2 {aJ_\nu(pz)\over bY(p\ep)} B_0C_0
  \left({z\over R} \right)^{-1-c_L+c_R}e^{-(m_L-m_R)z}
\end{equation}
and after taking $\ep \to 0$ and simplifying we find
\begin{equation}
  \label{eq:3}
  \Gamma_Y = -{\gamma_5\over \sqrt{2}} \sqrt{a\over b}
  R^{c_L-c_R-{3\over 2}}\sqrt{\pi} p^\nu b_0 c_0
  \int z^{1-c_L-c_R}e^{-(m_L-m_R)z} J_\nu(pz) dz
\end{equation}
where $b_0$ and $c_0$ are given in Eqs. \gl{eq:1a}--\gl{eq:2}. With
an eye towards our later definition of the Yukawa coupling in the 4D
theory, we redefine the conformal Yukawa coupling in Eq.~\gl{eq:3} as
\begin{equation}
  \label{eq:david3.27}
  \Gamma_Y= - {\lambda\over \Lambda^{d-1}}F(q^2)
\end{equation}
where $F(p^2)$ encodes all of the momentum dependence and $\lambda$ is
a dimensionless coupling constant. To find $F(p^2)$, we must evaluate
the integral over $z$ in Eq. \gl{eq:3}. It is given in \cite{david14}:
\begin{equation}
  \int_0^\infty e^{-at}J_\nu(bt)t^{\mu-1}dt = 
  \frac{ {1\over 2}b^\nu \Gamma(\mu+\nu) }{(a^2+b^2)^{{1\over 2}
      (\mu+\nu)}
    \Gamma(\nu+1)} {_2F_1}\left({\mu+\nu\over 2},{1-\mu+\nu\over 2}; \nu+1; {b^2\over b^2+a^2}
  \right)\,.
\end{equation}
The above integral is convergent if the following conditions are met:
${\rm{Re}}(a + ib) > 0$ and ${\rm{Re}}(a-ib)>0$ as well as
${\rm{Re}}(\nu+\mu)>0$. In our case, $b=p$, $a=m_L-m_R$ and
$\mu=2-c_L+c_R$. Thus we see that for convergence $c_L -c_R <
2+\nu$. Also, since $p$ is always real, the only other condition is
$m_L - m_R > 0$, which we already assumed to be true. Note that the
limits of the above integral are zero to infinity as opposed to
ranging from $\ep$ to infinity, which is the range of our integral
over $z$ in Eq.~\gl{eq:3}. This only introduces corrections of order
$\ep$, which will go to zero as we take $\ep\to 0$. The Yukawa
coupling then becomes
\begin{eqnarray}
  \nonumber
  \Gamma_Y= - {\lambda\over \sqrt{2}} \sqrt{{a\over b}}
  R^{c_L-c_R-1}\sqrt{\pi}
  p^\nu b_0c_0 \frac{ ({1\over 2}p)^\nu \Gamma(2-c_L+c_R+\nu)}
  { ( (m_L-m_R)^2+p^2)^{{1\over 2}(2-c_L+c_R+\nu}\Gamma(\nu+1)}\\
  \times\,
  {_2F_1}\left( {2-c_L+c_R+\nu\over 2}, {c_L-c_R+\nu-1\over 2}; 
    \nu+1; {p^2\over (m_L-m_R)^2+p^2} \right)
  \,.
\end{eqnarray}

\subsubsection*{The conformal Yukawa coupling at high momentum}
Since, to investigate unitarity, we are interested in the high energy
behavior of the theory, we will want to take the high $p$ limit of the
conformal Yukawa coupling. In the limit $p\gg m_L-m_R$ the last
argument of the hypergeometric function goes to one, allowing for
considerable simplification. Substituting $\nu = 2 - d_S$, we get
\begin{equation}
  \lim_{p\rig \infty} \Gamma_Y
  =  - {\lambda\over \sqrt{2}} \sqrt{{a\over b}} {\pi\over 2} b_0c_0 {\Gamma(2-c_L+c_R+\nu) \over
    \Gamma\left( {c_L-c_R+\nu\over 2}\right ) \Gamma\left( {3-c_L+c_R+\nu\over 2}\right ) 
  }
  (pR)^{c_L-c_R-1}p^{\nu-1}
\end{equation}

\subsubsection*{Making the fermions elementary}
In the Unhiggs theory, we take the fermions to be elementary. Setting
$c_L$ and $c_R$ to $1/2$ and $-1/2$ respectively yields a scaling
dimension for the fermions of $3/2$, allowing for their interpretation
as elementary. This yields the following form of the Yukawa coupling:
\begin{equation}
  \label{eq:david3.32}
  \lim_{p\rig \infty,c_L=1/2,c_R=-1/2} \Gamma_Y =  
  - {\lambda\over \sqrt{2}} \sqrt{{a\over b}} {\pi\over 2} b_0c_0 
  {\Gamma(3-d)\over \Gamma({3-d\over 2}) \Gamma({4-d\over 2})} p^{1-d}\,.
\end{equation}
From now on we will take the Yukawa coupling with elementary fermions
to be the definition of the Yukawa coupling in the Unhiggs theory. The
important conclusion is that the momentum dependence of the coupling
is simply $p^{1-d}$, so that
\begin{equation}
  \lim_{p\to \infty} F(p^2)=A_dp^{1-d}
\end{equation}
where $A_d$ is a momentum-independent constant defined by Eqs.~\gl{eq:david3.27} 
and \gl{eq:david3.32}. We find then that the
conformal coupling, unlike the non-conformal coupling, is a derivative
coupling, depending on the momentum scale to a non-integer power.

\subsection{Unitarity of $\bar t t \to W^+_L W^-_L$ with the conformal
  coupling}
Our previous calculation shows that the conformal Yukawa term in the
Lagrangian is now given in momentum space by
\begin{equation}
  \label{eq:david4.34}
  L_{Y,C}=-\lambda_t \bar t_R {H^\dagger(p)\over \Lambda^{d-1}}F(p^2) \left(
    \begin{matrix} t\\ b\end{matrix} \right)_L + \rm{h.c.}
\end{equation}
where $F(p^2)$ is given by $A_dp^{1-d}$ at high momentum.  The mass
term for the top quark is given by
\begin{equation}
  m_t={\lambda_tv^dF(0)\over \sqrt{2}\Lambda^{d-1}}\,.
\end{equation}
The value of $F(0)$ is arbitrary and for convenience we set it equal
to 1, so that the mass term takes the same form as in the case with
the non-conformal coupling (Eq. \gl{eq:david2.10}). The Yukawa
coupling between the physical Unhiggs and the $\bar t t$ pair is now
given by
\begin{equation}
  \Gamma_{Y,C}=i{m_t\over v^d} F(p^2)\,.
\end{equation}
Therefore the Unhiggs exchange diagram will have a factor of $F(p^2)$
which was not present in the case of the non-conformal coupling.

\begin{figure}
  \begin{center}
    \includegraphics[width=4cm]{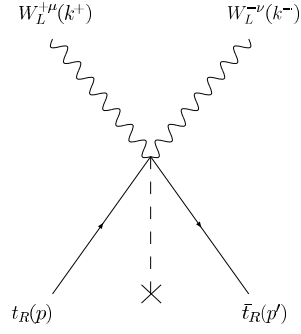}
  \end{center}
  \caption{\label{fig:david3} The contact diagram which results from
    gauging the non-local Yukawa coupling. The cross denotes that the
    Unhiggs is taken to its VEV.}
\end{figure}

There is another important consequence of the momentum dependence of
the conformal Yukawa term in Eq. \gl{eq:david4.34}. Because it
contains a derivative in position space, it must be gauged. Since it
is non-local, it will yield terms in the Lagrangian coupling the
Unhiggs, $\bar t t$ and arbitrary numbers of gauge bosons. More
specifically, there will be a term in the Lagrangian coupling the
Unhiggs, $\bar t t$ and two $W$ bosons. Upon taking the Unhiggs to its
VEV, this term will yield a contact interaction which will contribute
to our scattering process $\bar t t \to W^+W^-$. This diagram is shown
in figure \ref{fig:david3}. We now proceed to derive the Feynman rule
for this diagram. Using the result in \cite{david13}, we find that
after gauging Eq. \gl{eq:david4.34}, there is a term in the Lagrangian
given by
\begin{equation} {\cal{L}}_{Y,Gauge} =-g^2 {\lambda \over
    \Lambda^{d-1}} \bar t_R
  H^\dagger(p)T^aT^bA^a_\alpha(q_1)A^b_\beta(q_2)\left(\begin{matrix}
      t \\ b \end{matrix} \right)_L G^{\alpha\beta}(p,q_1,q_2)+
  {\rm{h.c.}}.
\end{equation}
where $T^a,T^b$ are the generators for the gauge fields $A^a_\alpha$,
$A^b_\beta$ and the function $G^{\alpha\beta}(p,q_1,q_2)$ is given by
\begin{eqnarray}
  G^{\alpha\beta}(p,q_1,q_2)=
  {g^{\alpha\beta}\over 2p\cdot (q_1+q_2)+(q_1+q_2)^2} \left[ F((p+q_1 +q_2)^2 )-F(p^2 )\right]\\
  +{{(2p+q_2)^\beta [2(p+q_2)+q_1]^\alpha\over 2(p+q_2)\cdot q_1+q_1^2}}
  \left[ {F((p+q_1+q_2)^2)-F(p^2)\over 2p\cdot(q_1+q_2)+(q_1+q_2)^2}
    -{F((p+q_2)^2)-F(p^2)\over 2p\cdot q_2+q_2^2}
  \right]
\end{eqnarray}
Upon identifying the gauge bosons with $W^+_\mu$ and $W^-_\nu$ (with
momenta $k^+$ and $k^-$, respectively), this leads to a contact
interaction between the top, anti-top and the two $W$ bosons given by
\begin{equation}
  \label{eq:david4.39}
  ig^2\Gamma_{Cont}^{+-\mu\nu} = i{g^2\over 2} {\lambda_t \over \Lambda^{d-1}} {v^d\over \sqrt{2}} 
  G^{\mu\nu}(0,k^+ ,k^- )= {-ig^2m_t\over 2}G^{\mu\nu}(0,k^+,k^-)\,.
\end{equation}

%%%%%%%%%%%%%%%%%%%%%%%%%%%%%%%%%%%%%%%%%%%%%%%%
%%%%%%%%%%%%%%%%%%%%%%%%%%%%%%%%%%%%%%%%%%%%%%%%

We now have two ``Unhiggs'' diagrams to consider: the first being the
conformal Unhiggs $s$-channel exchange diagram in figure
\ref{fig:david2} and the second being the contact diagram in figure
\ref{fig:david3}. The exchange diagram amplitude is simply given by
the non-conformal amplitude from Eq.~\gl{eq:david2.15} multiplied by
$F(s)$.
\begin{equation}
  \label{eq:david4.40}
  {\cal{M}}_{h,C}=-\sqrt{2} G_F m_t \sqrt{s} F(s) \left[ 
    {\mu^{4-2d}-(\mu^2-s)^{2-d}\over \mu^{4-2d} -m^{4-2d} -(\mu^2-s)^{2-d}} 
  \right]
\end{equation}
The contribution to the amplitude from the contact diagram is given by
\begin{equation} 
  {\cal{M}}_{Cont} = \bar v(p')g^2
  \Gamma^{+-\mu\nu}_{Cont} \ep_\mu(k^+)\ep_\nu(k^-)
\end{equation}
which simplifies once we contract the vertex in Eq.~\gl{eq:david4.39}
with the $W$ polarization vectors. This yields
\begin{equation}
  \ep_\mu(k^+)\ep_\nu(k^-) G^{\mu\nu}(0,k^+,k^-) = 
  \ep(k^+)\cdot \ep(k^-) {F(k^+ + k^-)-F(0) \over (k^++k^-)^2} \,.
\end{equation}
Finally, putting the contributions from Eqs. \gl{eq:david2.6},
\gl{eq:david4.40} and \gl{eq:david4.43} together, the tree level
amplitude for the process $\bar t t \to W_L^+ W_L^-$ is given by
\begin{eqnarray}
  \label{eq:david4.43}
  {\cal{M}}_C =& \sqrt{2}G_F  m_t \sqrt{s}+ \sqrt{2}G_F m_t \sqrt{s}[F(s)-1] 
  - \sqrt{2} G_F m_t \sqrt{s} F(s) \left[
    {\mu^{4-2d}-(\mu^2-s)^{2-d}\over \mu^{4-2d} -m^{4-2d}
      -(\mu^2-s)^{2-d}} \right] \,.
\end{eqnarray}
Upon taking the limit $s\gg \mu , m^2$ this yields
\begin{equation} 
  {\cal{M}} = \sqrt{2}G_F m_t m^{4-2d} (-1)^{d-2}
  F(s)s^{d- {3\over 2}}
\end{equation}
The only difference between this amplitude and the non-conformal
amplitude is the presence of $F(s)$. This is a crucial difference,
however, as upon taking the high energy limit of $F(s)$ we find that
the amplitude becomes
\begin{equation}
  {\cal{M}}_C= \sqrt{2}G_F m_t m^{4-2d}(-1)^{d-2}A_d
  s^{1-d\over 2} s^{d-{3\over 2}} = \sqrt{2}G_F m_t m^{4-2d}
  (-1)^{d-2}A_d s^{d-2\over 2}
\end{equation}
We see that for all $d$ in the range $1 \leq d < 2$ the amplitude is a
decreasing function of $s$, and therefore the process does not violate
unitarity.

\section{Summary \& Conclusion}
The Unhiggs scenario \cite{david6} has interesting phenomenological
implications, which can be tested at the LHC, already with early
data \cite{zzchannel,prep}.

While spontaneous symmetry breaking assures that the weak boson fusion
channels with clean leptonic final states remain unaltered, the
unitarity bounds that result from the interplay of the Yukawa and
gauge sector leave constraints on the model and imply a modified
phenomenology of gluon fusion channels. We have
investigated these modifications in the clean gold-plated mode $H\rig
ZZ \rig 4~{\rm{leptons}}$. Measurements in this clean channel will
provide stringent bounds on the Unhiggs scenario, with only a minimum
of experimental uncertainties. On the one hand, if a SM Higgs
candidate in the mass range where this channel is important is
observed in the future, we show that the Unhiggs hypothesis can in
principle be tested at the percent-level in this channel. On the other
hand the non-observation of Higgs can be accomodated in the Unhiggs
scenario with $d\simeq 1.1$.

For $d>1.5$ the low energy effective theory has to be completed by a
conformally invariant Yukawa coupling to avoid unitarity violation in
the massive quark scattering $\bar t t \rig VV$ amplitudes. We achieve
this by extending the $\bar t tH$ coupling of Ref.~\cite{david6} to a
``conformal'' Yukawa coupling. Within the framework of AdS/CFT we
demonstrate that this modified coupling serves to restore unitarity
over the full parameter range $1\leq d < 2$ in the high energy limit.
This is tantamount to a modification of the heavy fermion sectors and
an extended fermionic spectrum, and we leave a phenomenological
analysis of these implications to future work.

\section*{Acknowledgements}
C.E. and M.S. thank the physics department of the University of
California, Davis, for hospitality.  C.E. acknowledges funding by the
Durham International Junior Research Fellowship scheme.  J.T. was
supported by the Department of Energy under grant DE-FG02-91ER406746.

%--------------------------------------------------------------------
%			BIBLIOGRAPHY
%--------------------------------------------------------------------

\end{document}